\documentclass[11 pt,oneside,onecolumn,a4paper]{article}
\usepackage{amsmath}
\usepackage{graphicx}
\DeclareGraphicsExtensions{.eps,.ps,.pdf}
\usepackage{bbm}
\usepackage{bm}
\usepackage{epsfig}
\usepackage[T1]{fontenc}
\usepackage{esint}
\usepackage{amssymb}

\usepackage{lipsum}
\usepackage{savesym}

mm \leftmargin 5pt \rightmargin 5pt \hoffset= -15mm

\title{Bath correlation functions for logarithmic spectral densities}
\author{Filippo Giraldi}

\date{\small{School of Chemistry and Physics, University of KwaZulu-Natal\\ 
and National Institute for Theoretical Physics (NITheP)\\
Westville Campus, Durban 4000, South Africa
\vspace{1em}\\Gruppo Nazionale per la
Fisica Matematica (GNFM-INdAM)\\
c/o Istituto Nazionale di Alta Matematica Francesco Severi\\
Citt\'a Universitaria, Piazza Aldo Moro 5, 00185 Roma, Italy}}

\begin{document}

\maketitle

\def\bbm[#1]{\mbox{\boldmath$#1$}}

\vspace{0em}

\vspace{0em}

 \begin{abstract}
 We study the bath correlation functions (BCFs) of open quantum systems interacting with thermal baths, in case the spectral densities (SDs) exhibit removable logarithmic singularities at low frequencies and are arbitrarily shaped at higher frequencies. The singularities consist in arbitrarily positive or negative powers of logarithmic functions, as additional factors for the power laws of the Ohmic-like SDs. If the SD vanishes sufficiently fast at high frequencies 
the short time behavior of the BCF is algebraic. The long time behavior of the BCF exhibits a variety of relaxations that involve inverse power laws and arbitrary powers of logarithmic forms.
The imaginary part of the BCF shows over long times regular dependence on the low frequency structure of the SD, except for certain conditions where the ohmicity parameter takes odd natural values. Same dependence holds for the real part of the BCF at non-vanishing temperatures. At zero temperature the real part of the BCF exhibits over long times the same regular relationship with the low frequency structure of the SD, except for certain conditions involving even natural values of the ohmicity parameter. The exceptional conditions provide relaxations that are faster than those obtained via the regular dependence. In this way, various long time relaxations of the BCF that are slower than exponential decays and arbitrarily faster or slower than inverse power laws, can be interpreted in terms of removable logarithmic singularities in the low frequency structure SD of an open quantum system.
\end{abstract}

 \maketitle

\section{Introduction}
\vspace{-0em}
The theory of open quantum systems allows a description of the open dynamics in terms of the features of the external environment \cite{BP,Weiss}. Usually, the environment is modeled by a collection of field modes or harmonic oscillators. The interaction between the open system and the external environment consists in a frequency dependent linear coupling between each mode and the degrees of freedom of the open system. The open dynamic is fully characterized by the spectral density (SD) of the system.  For the sake of clarity we report below a brief description of these quantities and refer to \cite{BP,Weiss,MS} for details. 

The total Hamiltomian $H$ of the open system and the external environment is usually described by the following form, $H=H_S+H_{SE}+H_E$, where $H_S$ is the microscopic Hamiltonian of the open system that acts on the corresponding Hilbert space $\mathcal{H}_S$ \cite{BP,Weiss}. The Hamiltonian $H_E$ acts on the Hilbert space $\mathcal{H}_E$ of the external environment and is represented as follows, $H_E=\sum_k \omega_k b^{\dagger}_k b_k$. The rising operator $b^{\dagger}_k$ and the lowering operator $b_k$ act on the Hilbert space of the $k$th mode or oscillator of frequency $\omega_k$. If the interaction between the open system and external environment is linear, the term $H_{SE}$ usually takes the form $H_{SE}=\sum_k\left(g_k L\otimes b^{\dagger}_k+g^{\dagger}_k L^{\dagger}\otimes b_k\right)$. The operator $L$ acts on the Hilbert space $\mathcal{H}_S$ of the open system and $g_k$ represents the coupling constant between the open system and 
the $k$th frequency mode \cite{HuPRB2002,RitEisfSDJCP2014,MS}. The spectral density (SD) of the system $J\left(\omega\right)$ is defined via the coupling constants as follows,
\begin{eqnarray}
J\left(\omega\right)= \sum_k \left|g_k\right|^2 \delta \left(\omega-\omega_k\right).\label{SD}
\end{eqnarray}
 This function encodes information about the environment, via the mode frequencies, and about the interaction between open system and environment, via the coupling constants. 

The open dynamics is influenced by the temporal correlations among the degrees of freedom of the environment. These effects are stored in the bath correlation (BCF). If the external environment is a thermal bath at temperature $T$, if no correlation exists initially between the open system and the thermal bath, and if the stationarity condition holds, the BCF is expressed in terms of the SD by the following form,
\begin{eqnarray}
&&C(t)=\int_0^{\infty} J\left(\omega\right)\left(
\cos\left(\omega t\right) \coth\frac{\omega}{2 T}-\imath \sin\left(\omega t\right)\right)\, d\omega,  \label{CT} 
\end{eqnarray}
in units where the Planck constant $\hbar$ and Boltzmann constant $k_B$ equal unity, $\hbar=k_B=1$. We refer to \cite{BP,Weiss,MS,RitEisfSDJCP2014,BCF_EisfeldPRA2015} for details.

In Ref. \cite{RitEisfSDJCP2014} a powerful technique has been developed for the analytical evaluation of the BCF by fitting the SD via a class of simple functions. In this way, the BCF is represented as a sum of damped harmonic oscillations. This form is useful for the description of the reduced dynamics of open quantum systems. The method allows the evaluation of the BCF for a large variety of SDs that include the ohmic and super-ohmic condition and forms appearing in models of photosynthetic light harvesting complexes, to name a few. 

As a continuation of the scenario described above, here, we consider a general class of SDs that exhibit at low frequencies logarithmic factors in addition to the ohmic-like structure. These terms are represented by arbitrarily positive or negative powers of logarithmic functions. The corresponding logarithmic singularities are removable and provide SDs that are arbitrarily shaped at higher frequencies. We analyze the short and long time behavior of the BCF in dependence on the low and high frequency structured of these SDs, that include and continuously depart from the ohmic-like condition.

The paper is organized as follows. Section \ref{2} is devoted
to the definition of the SDs under study. In Section \ref{3}, the short and long time behavior of the BCF is evaluated analytically in terms of the low or high frequency structure and integral properties of the SD. Conclusions are drawn in Section \ref{5} and details of the calculations are provided in the Appendix.

\section{Logarithmic spectral densities 
}\label{2}

For the sake of convenience, the SDs under study are described via the dimensionless auxiliary function $\Omega\left(\nu\right)$. This function is defined for every $\nu\geq0$ by the following scaling property, $\Omega\left(\nu \right)=J \left( \omega_s\nu \right)/ \omega_s $, where $\omega_s$ is a typical scale frequency of the system. The first class of SDs under study is defined by auxiliary functions 
that are continuous for every $\nu> 0$ and exhibit the following asymptotic behavior \cite{BleisteinBook} as $\nu\to 0^+$,
\begin{eqnarray}
&&\hspace{-0em}\Omega\left(\nu\right)\sim 
\sum_{j=0}^{\infty}
\sum_{k=0}^{n_j}c_{j,k} \nu^{\alpha_j}\left(- \ln \nu\right)^k,  \label{o0log} 
\end{eqnarray}
 where $\alpha_0>0$, $\infty> n_j\geq 0$, $\alpha_{j+1}>\alpha_j$ for every $j\geq 0$, and $\alpha_j\uparrow +\infty$ as $j\to +\infty$. Notice that ohmic-like SDs \cite{BP,Weiss} are obtained as a special case for $n_0=0$. For this reason, the power $\alpha_0$ is referred as the ohmicity parameter. If $n_0=0$, the corresponding SDs are super-ohmic as $\omega\to 0^+$ for $\alpha_0>1$, ohmic for $\alpha_0=1$ and sub-ohmic for $0<\alpha_0<1$. The logarithmic singularity in $\nu=0$ is removed by defining $\Omega(0)=0$. The summability of the SD is guaranteed by requiring as $\nu\to+\infty$ the following asymptotic behaviors, $\Omega\left(\nu\right)= O\left(\nu^{-1-\chi_0}\right)$, where $\chi_0>0$. Let the complex valued functions $\hat{\Omega}\left(s\right)$ and $\hat{\Omega}_T\left(s\right)$ represent the Mellin transform \cite{BleisteinBook,Wong-BOOK1989} of the auxiliary function $\Omega\left(\nu\right)$ and of the function $\Omega_T\left(\nu\right)$, that is defined as follows, $\Omega_T\left(\nu\right)=\Omega\left(\nu\right) \coth \left(\omega_s \nu/\left(2 T\right)\right)$. The functions $\hat{\Omega}\left(s\right)$ and $\hat{\Omega}_T\left(s\right)$ and the meromorphic continuations \cite{BleisteinBook,Wong-BOOK1989} are required to decay sufficiently fast as $\left|\mathrm{Im} \,s \right|\to+\infty$. Details are provided in the Appendix.

In light of the asymptotic analysis performed in Ref. \cite{WangLinJMAA1978,Wong-BOOK1989}, the second class of SDs under study is described by auxiliary functions with the following asymptotic expansion as $\nu\to 0^+$, 
\begin{eqnarray}
&&\hspace{-0em}\Omega\left(\nu\right)\sim \sum_{j=0}^{\infty}w_j
\, \nu^{\alpha_j} \left(-\ln \nu\right)^{\beta_j}.  \label{OmegaLog0}
\end{eqnarray}
The powers $\alpha_j$ fulfill the constraints reported above, while the powers $\beta_j$ are considered to be just real valued, either positive, vanishing or negative. Again, the logarithmic singularity in $\nu=0$ is removed by setting $\Omega(0)=0$.
Let the parameter $\bar{n}$ be the least natural number such that $\alpha_{k-1}+1\leq \bar{n}<\alpha_{k}+1$, where the index $k$ is a non-vanishing natural number. The function $\Omega^{\left(\bar{n}\right)}\left(\nu\right)$ is required to be continuous on the interval $\left(0,\infty\right)$. The integral $\int_0^{\infty}\Omega\left(\nu\right)\exp\left\{-\imath \xi \nu\right\} d \nu$ must converge uniformly for all sufficiently large values of the variable $\xi$ and the integral $\int \Omega^{\left(\bar{n} \right)}\left(\nu\right)\exp\left\{-\imath \xi \nu\right\} d \nu$ has to converge at $\nu=+\infty$ uniformly for all sufficiently large values of the variable $\xi$. The auxiliary function 
is required to be differentiable $k$ times and to fulfill the following asymptotic expansion as $\nu\to 0^+$,
$$\Omega^{(k)}\left(\nu\right)\sim \sum_{j=0}^{\infty}w_j
\, \frac{d^k}{d\nu^k}\left(\nu^{\alpha_j} \left(-\ln \nu\right)^{\beta_j}\right),$$
for every $k=0,1, \ldots,\bar{n} $. Furthermore, for every $k=0, \ldots,\bar{n}-1$, the function $\Omega^{(k)}\left(\nu\right)$ has to vanish as $\nu\to +\infty$. See Ref. \cite{WangLinJMAA1978} for details.

If compared to the first class, the second class of SDs under study has to fulfill more constraints but includes arbitrary positive, vanishing or negative powers of logarithmic forms. In both the classes under study the auxiliary functions $\Omega\left(\nu\right)$ are non-negative, bounded and summable, due to physical grounds, and, apart from the above constraints, arbitrarily shaped. For the sake of shortness, we refer to the above classes as logarithmic SDs in order to highlight the presence of removable logarithmic singularities in the low frequency structure.

\section{The bath correlation function}\label{3}

The BCF, given by Eq. (\ref{CT}), is uniquely determined by the SD and the temperature of the thermal bath. The low frequency and integral properties of the SD characterize the long time behavior of the BCF. For the sake of simplicity, we define the functions $C_1(t)$ and $C_2(t)$ as the real part and the opposite of the imaginary part of the BCF, respectively, $C_{1,T}(t)= \mathrm{Re}\,\left\{C(t)\right\}$ and $C_2(t)=- \mathrm{Im}\,\left\{C(t)\right\}$, this means $C(t)=C_{1,T}(t)-\imath C_2(t)$. The functions $C_{1,T}(t)$ and $C_2(t)$ vanish over long times since the constraint 
\begin{eqnarray}
\int_0^{\infty}
 J\left(\omega\right)\,\coth \frac{ \omega}{2 T}\,d \omega<\infty, \label{cond1T}
\end{eqnarray}
holds for the SDs under study, and due to the summability of the SD.

Over short times, the behavior of BCF depends on the high frequency structure and integral properties of the SD. In fact, the function $C_{1,T}(t)$ departs from the finite value $C_{1,T}(0)$ in various ways that are determined by the decay of the SD at high frequencies. 
A simplified form is obtained for SDs that are tailored at high frequency, $\omega\gg\omega_s$, as follows, $J\left(\omega\right)=O\left(\left(\omega/\omega_s\right)^{-1-\chi_0}\right)$ with $\chi_0>2$. In this case the function $C_{1,T}(t)$ decreases quadratically for $t\ll 1/\omega_s$,
\begin{eqnarray}
C_{1,T}(t)\sim C_{1,T}(0)-l_{1,T} t^2, \label{C1Tt0}
\end{eqnarray}
where $C_{1,0}(0)=\int_0^{\infty}J\left(\omega\right)\,d \omega$, $l_{1,T}= \int_0^{\infty} \omega^2 J\left(\omega\right)\coth \left( \omega/\left(2 T\right)\right)\,d \omega/2$ for $T>0$, and $l_{1,0}= \int_0^{\infty} \omega^2 J\left(\omega\right)\,d \omega/2$. Similarly, over short times the function $C_2(t)$ exhibits various behaviors, depending on the high frequency decay of the SD. Consider high frequency profiles where  $J\left(\omega\right)=O\left(\left(\omega/\omega_s\right)^{-1-\chi_0}\right)$ for $\omega\gg\omega_s$ with $\chi_0>1$. In this case the function $C_2(t)$ grows linearly for $t\ll 1/\omega_s$,
\begin{eqnarray}
C_2(t)\sim l_2 t, \label{C2t0}
\end{eqnarray}
where $l_2= \int_0^{\infty} \omega J\left(\omega\right)\,d \omega$.

\subsection{Long time logarithmic and power law relaxations}

At non-vanishing temperatures, $T>0$, if the SDs belong to the first class under study, the function $C_{1,T}(t)$ vanishes over long times according to relaxations that involve logarithmic and power laws. If the ohmicity parameter is not an odd natural number we find for $t \gg 1/\omega_s$ the following logarithmic decays,
\begin{eqnarray}
C_{1,T}(t)\sim c_{0,n_0} r_{1,T} \left(\omega_s t\right)^{-\alpha_0} \ln^{n_0}\left(\omega_s t\right), \label{C1tInfT}
\end{eqnarray}
where $r_{1,T}=2 \omega_s T \cos \left(\pi \alpha_0/2\right) \Gamma\left(\alpha_0\right)$. The above form provides inverse power laws for $n_0=0$,
\begin{eqnarray}
C_{1,T}(t)\sim c_{0,n_0} r_{1,T} \left(\omega_s t\right)^{-\alpha_0}. \label{C1tInfTpl}
\end{eqnarray}
Faster logarithmic relaxations are obtained for $\alpha_0=1+2m_0$, where $m_0$ is a natural number and $n_0$ is a non-vanishing natural number,
\begin{eqnarray}
C_{1,T}(t)\sim c_{0,n_0}\bar{r}_{1,T} \left(\omega_s t\right)^{-1-2 m_0} \ln^{n_0-1}\left(\omega_s t\right). \label{C1tInfT1}
\end{eqnarray}
The parameter $\bar{r}_{1,T}$ reads $\bar{r}_{1,T}=\pi  \omega_s T n_0 (-1)^{m_0} \left(2 m_0\right)!\, $.
The above asymptotic expression gives inverse power laws for $n_0=1$,
\begin{eqnarray}
C_{1,T}(t)\sim c_{0,n_0}\bar{r}_{1,T} \left(\omega_s t\right)^{-1-2 m_0}. \label{C1tInfT1pl}
\end{eqnarray}
Even faster relaxations are found if $\alpha_0=1+2m_0$ and $n_0=0$. Consider the least positive index $k_0$ such that either $\alpha_{k_0}$ is not an odd natural number, 
or $\alpha_{k_0}= 1+2m_{k_0}$, where $m_{k_0}$ and $n_{k_0}$ are non-vanishing natural numbers. The long time behavior of the function $C_{1,T}(t)$ is obtained from Eqs. (\ref{C1tInfT}) and (\ref{C1tInfTpl}) by substituting the parameter $\alpha_0$ with $\alpha_{k_0}$ and $n_0$ with $n_{k_0}$, in the former case, or by replacing in Eqs. (\ref{C1tInfT1}) and (\ref{C1tInfT1pl}) the number $m_0$ with $m_{k_0}$ and $n_0$ with $n_{k_0}$, in the latter case. We consider SDs such that the index $k_0$ exists with the required properties.

If the SDs belong to the second class under study, the function $C_{1,T}(t)$ exhibits for $t \gg 1/\omega_s$ a variety of relaxations that involve arbitrarily positive, vanishing or negative powers of logarithmic forms,
\begin{eqnarray}
C_{1,T}(t)\sim w_0  \left(\omega_s t\right)^{-\alpha_0} \left(r_{1,T}\ln^{\beta_0}\left(\omega_s t\right)+\bar{\bar{r}}_{1,T}\ln^{\beta_0-1}\left(\omega_s t\right) \right), \label{C1tInfTSD2}
\end{eqnarray}
where $\bar{\bar{r}}_{1,T}=2 \omega_s T\left(\pi \beta_0 \sin\left(\pi \alpha_0/2\right) \Gamma\left(\alpha_0\right)/2-\beta_0 \cos\left(\pi \alpha_0/2\right)\Gamma^{(1)}\left(\alpha_0\right)\right)$. If the ohmicity parameter does not take odd natural values, the dominant part of the above relaxation is $C_{1,T}(t)\sim w_0 r_{1,T} \left(\omega_s t\right)^{-\alpha_0} \ln^{\beta_0}\left(\omega_s t\right)$, and turns into the inverse power laws $C_{1,T}(t)\sim w_0 r_{1,T} \left(\omega_s t\right)^{-\alpha_0}$ for $\beta_0=0$. If the ohmicity parameter is an odd natural number, Eq. (\ref{C1tInfTSD2}) provides faster logarithmic relaxations, $C_{1,T}(t)\sim w_0 \bar{\bar{r}}_{1,T}\left(\omega_s t\right)^{-\alpha_0} \ln^{\beta_0-1}\left(\omega_s t\right)$, that become the inverse power laws $C_{1,T}(t)\sim w_0 \bar{\bar{r}}_{1,T} \left(\omega_s t\right)^{-\alpha_0}$ for $\beta_0=1$.

The function $C_{1,0}(t)$ vanishes over long times according to various relaxations that involve logarithmic and power laws for SDs belonging to the first class under study. If the ohmicity parameter is not an even natural number we find for $t \gg 1/\omega_s$ the following form,
\begin{eqnarray}
C_{1,0}(t)\sim -c_{0,n_0} r_{1,0} \left(\omega_s t\right)^{-1-\alpha_0} \ln^{n_0}\left(\omega_s t\right), \label{C1tInf0}
\end{eqnarray}
where $r_{1,0}= \omega^2_s \sin\left(\pi \alpha_0/2\right) \Gamma\left(1+\alpha_0\right)$. The above relaxations become inverse power laws for $n_0=0$,
\begin{eqnarray}
C_{1,0}(t)\sim -c_{0,n_0} r_{1,0} \left(\omega_s t\right)^{-1-\alpha_0}. \label{C1tInf0pl}
\end{eqnarray}
Faster logarithmic decays appear for $t \gg 1/\omega_s$ if $\alpha_0= 2m_1$, where $m_1$ and $n_0$ are non-vanishing natural numbers,
\begin{eqnarray}
C_{1,0}(t)\sim c_{0,n_0} \bar{r}_{1,0} \left(\omega_s t\right)^{-1-2 m_1} \ln^{n_0-1}\left(\omega_s t\right). \label{C1tInf0m1}
\end{eqnarray}
The parameter $\bar{r}_{1,0}$ reads $\bar{r}_{1,0}=\pi \omega_s^2n_0(-1)^{m_1}\left(2 m_1\right)!\, /2$. The above asymptotic relaxations become inverse power laws for $n_0=1$,
\begin{eqnarray}
C_{1,0}(t)\sim c_{0,n_0} \bar{r}_{1,0} \left(\omega_s t\right)^{-1-2 m_1}. \label{C1tInf0m1pl}
\end{eqnarray}
Even faster relaxations are found if $\alpha_0= 2m_1$ and $n_0=0$. Consider the least non-vanishing index $k_1$ such that either $\alpha_{k_1}$ is not an even natural number or $\alpha_{k_1}= 2m_{k_1}$, where $m_{k_1}$ and $n_{k_1}$ are non-vanishing natural numbers. The function $C_{1,0}(t)$ is obtained, in the former case, from Eqs. (\ref{C1tInf0}) and (\ref{C1tInf0pl}) by substituting the ohmicity parameter with $\alpha_{k_1}$ and $n_0$ with $n_{k_1}$, and, in the latter case, from Eqs. (\ref{C1tInf0m1}) and (\ref{C1tInf0m1pl}) by substituting the parameter $m_1$ with $m_{k_1}$ and $n_0$ with $n_{k_1}$.

If the SDs belong to the second class under study, we obtain for $t \gg 1/\omega_s$ a variety of relaxations that involve arbitrary positive or negative powers of logarithmic forms,
\begin{eqnarray}
C_{1,0}(t)\sim w_0\left(\omega_s t\right)^{-1-\alpha_0} 
\left( - r_{1,0}  \ln^{\beta_0} \left(\omega_s t\right) + \bar{\bar{r}}_{1,0}  \ln^{\beta_0-1} \left(\omega_s t\right)\right), \label{C1tInf0SD2}
\end{eqnarray}
where $\bar{\bar{r}}_{1,0}=\pi \beta_0\Gamma\left(1+\alpha_0\right)\cos\left(\pi \alpha_0/2\right)/2+\beta_0 \sin\left(\pi \alpha_0/2\right)\Gamma^{(1)}\left(1+\alpha_0\right)$. If the ohmicity parameter does not take even natural values, the dominant part of 
the above asymptotic form is $C_{1,0}(t)\sim - w_0 r_{1,0} \left(\omega_s t\right)
^{-1-\alpha_0} \ln^{\beta_0}\left(\omega_s t\right)$, and gives the inverse power laws $C_{1,0}(t)\sim -w_0 r_{1,0}
 \left(\omega_s t\right)^{-1-\alpha_0}$ for $\beta_0=0$. If the ohmicity parameter is an even natural number, Eq. (\ref{C1tInf0SD2}) provides faster logarithmic relaxations,
$C_{1,0}(t)\sim w_0 \bar{\bar{r}}_{1,0}\left(\omega_s t\right)
^{-1-\alpha_0} \ln^{\beta_0-1}\left(\omega_s t\right)$, that result in the inverse power laws $C_{1,0}(t)\sim w_0 \bar{\bar{r}}_{1,0} \left(\omega_s t\right)^{-\alpha_0}$ for $\beta_0=1$.

If the ohmicity parameter is not an odd natural number, the function $C_{2}(t)$ vanishes for $t \gg 1/\omega_s$ as below,
\begin{eqnarray}
C_{2}(t)\sim c_{0,n_0} r_{2} \left(\omega_s t\right)^{-1-\alpha_0} \ln^{n_0}\left(\omega_s t\right), \label{C2tInf0}
\end{eqnarray}
where $r_{2}=\omega^2_s \cos \left(\pi \alpha_0/2\right) \Gamma\left(1+\alpha_0\right)$.
The above expression describes inverse power laws for $n_0=0$,
\begin{eqnarray}
C_{2}(t)\sim c_{0,n_0} r_{2} \left(\omega_s t\right)^{-1-\alpha_0}. \label{C2tInf0pl}
\end{eqnarray}
We find faster logarithmic relaxations for $t \gg 1/\omega_s$ if $\alpha_0=1+2m_2$, where $m_2$ is a natural number and $n_0$ is a non-vanishing natural number,
\begin{eqnarray}
C_{2}(t)\sim c_{0,n_0} \bar{r}_{2} \left(\omega_s t\right)^{-2-2 m_2} \ln^{n_0-1}\left(\omega_s t\right). \label{C2tInfm2}
\end{eqnarray}
The parameter $\bar{r}_{2}$ reads $\bar{r}_{2}=\pi n_0 \omega_s^2(-1)^{m_2} \left(1+2 m_2\right)! /2$.
The above asymptotic form provides inverse power laws for $n_0=1$,
\begin{eqnarray}
C_{2}(t)\sim c_{0,n_0} \bar{r}_{2} \left(\omega_s t\right)^{-2-2 m_2}. \label{C2tInfm2pl}
\end{eqnarray}
Even faster relaxations are obtained if $\alpha_0 =1+2 m_2$ and $n_0=0$. Consider the least non-vanishing index $k_2$ such that either $\alpha_{k_2}$ is not an odd natural number or $\alpha_{k_2}=1+ 2m_{k_2}$, where $m_{k_2}$ is a natural number, and $n_{k_2}$ is a non-vanishing natural number. The function $C_{2}(t)$ is obtained, in the former case, from Eqs. (\ref{C2tInf0}) and (\ref{C2tInf0pl}) by substituting the parameter $\alpha_0$ with $\alpha_{k_2}$ and $n_0$ with $n_{k_2}$, and, in the latter case, from Eqs. (\ref{C2tInfm2}) and (\ref{C2tInfm2pl}) by substituting the parameter $m_2$ with $m_{k_2}$ and $n_0$ with $n_{k_2}$.

For the second class of SDs under study, we obtain for $t \gg 1/\omega_s$ a variety of relaxations that involve arbitrarily positive, vanishing or negative powers of logarithmic forms,
\begin{eqnarray}
C_{2}(t)\sim w_0 \left(\omega_s t\right)^{-1-\alpha_0}
\left(r_{2}  \ln^{\beta_0}\left(\omega_s t\right)+
\bar{\bar{r}}_{2}  \ln^{\beta_0-1}\left(\omega_s t\right)
\right), \label{C2tInf0SD2}
\end{eqnarray}
where $\bar{\bar{r}}_{2}=\beta_0 \left(
\pi \sin\left(\pi \alpha_0/2\right) 
\Gamma\left(1+\alpha_0\right)/2-\cos\left(\pi \alpha_0/2\right)
\Gamma^{(1)}\left(1+\alpha_0\right)
 \right)$. If the ohmicity parameter does not take odd natural values, the dominant part of the above asymptotic form is $C_{2}(t)\sim w_0 r_{2} \left(\omega_s t\right)
^{-1-\alpha_0} \ln^{\beta_0}\left(\omega_s t\right)$, and provides the inverse power laws $C_{2}(t)\sim w_0 r_{2}
 \left(\omega_s t\right)^{-1-\alpha_0}$ for $\beta_0=0$. If the ohmicity parameter is an odd natural number, Eq. (\ref{C2tInf0SD2}) gives faster logarithmic relaxations,
$C_{2}(t)\sim w_0 \bar{\bar{r}}_{2}\left(\omega_s t\right)
^{-1-\alpha_0} \ln^{\beta_0-1}\left(\omega_s t\right)$, that turn into the inverse power laws $C_{2}(t)\sim w_0 
\bar{\bar{r}}_{2} \left(\omega_s t\right)^{-1-\alpha_0}$ for $\beta_0=1$.

\begin{figure}[t]
\centering
\includegraphics[height=6.25 cm, width=10.25 cm]{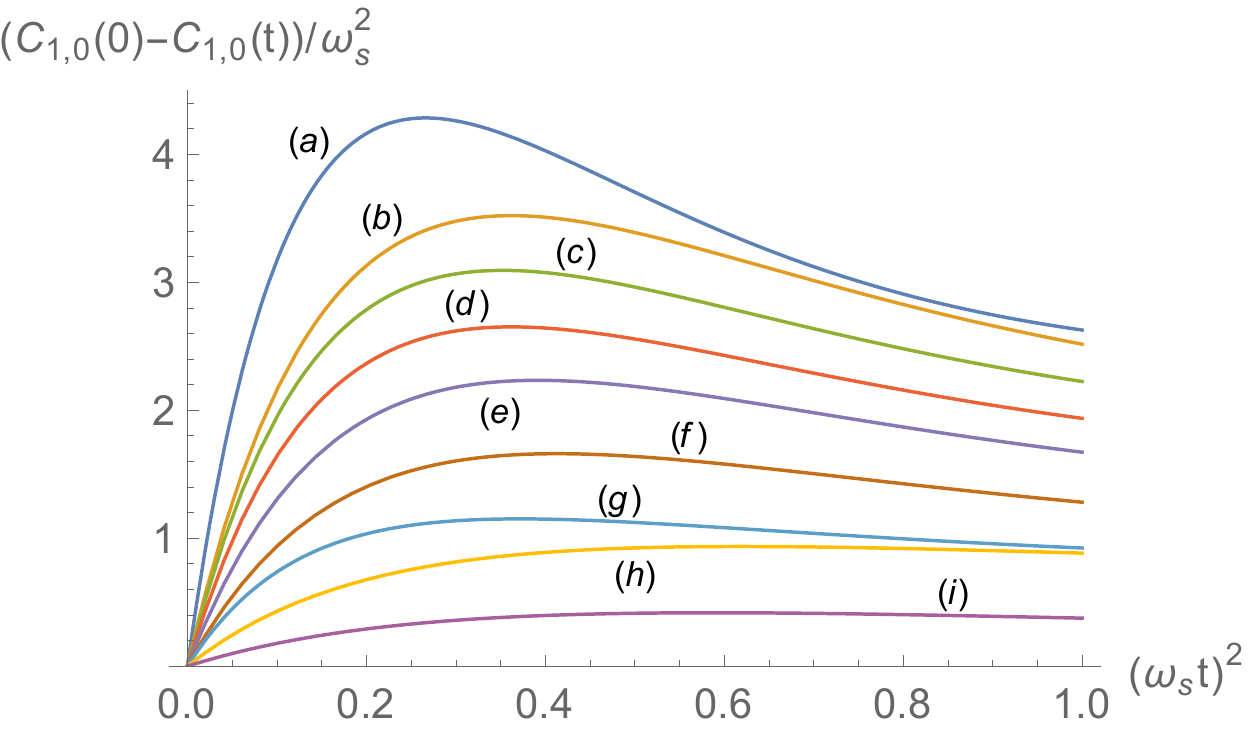}
\vspace{0.2cm}
\caption{(Color online) The quantity 
$\left(C_{1,0}(t)-C_{1,0}(0)\right)/\omega_s^2 $ versus $\left( \omega_s t\right)^2$ for $0\leq\omega_s t\leq 1$, $J\left(\omega\right)=q\omega_s \left(\omega/\omega_s\right)^{\alpha}\exp\left\{- l \omega/\omega_s \right\}\left|\ln\left(\omega/\omega_s \right)\right|^n$, $q=1$ and different values of the parameters $\alpha$, $l$ and $n$. The curve $(a)$ corresponds to $\alpha=1$, $l=1$ and $n=6$; $(b)$ corresponds to $\alpha=2.1$, $l=1.2$ and $n=4$, $(c)$ corresponds to $\alpha=1.7$, $l=1.1$ and $n=4$; $(d)$ corresponds to $\alpha=1.6$, $l=1.1$ and $n=4$; $(e)$ corresponds to $\alpha=1.8$, $l=1.2$ and $n=4$; $(f)$ corresponds to $\alpha=1.6$, $l=1.2$ and $n=4$; $(g)$ corresponds to $\alpha=0.1$, $l=0.8$ and $n=4$; $(h)$ corresponds to $\alpha=1$, $l=1$ and $2$; $(i)$ corresponds to $\alpha=1.5$, $l=1.5$ and $n=4$.}
\label{Fig1SD}
\end{figure}

\begin{figure}[t]
\centering
\includegraphics[height=6.25 cm, width=10.25 cm]{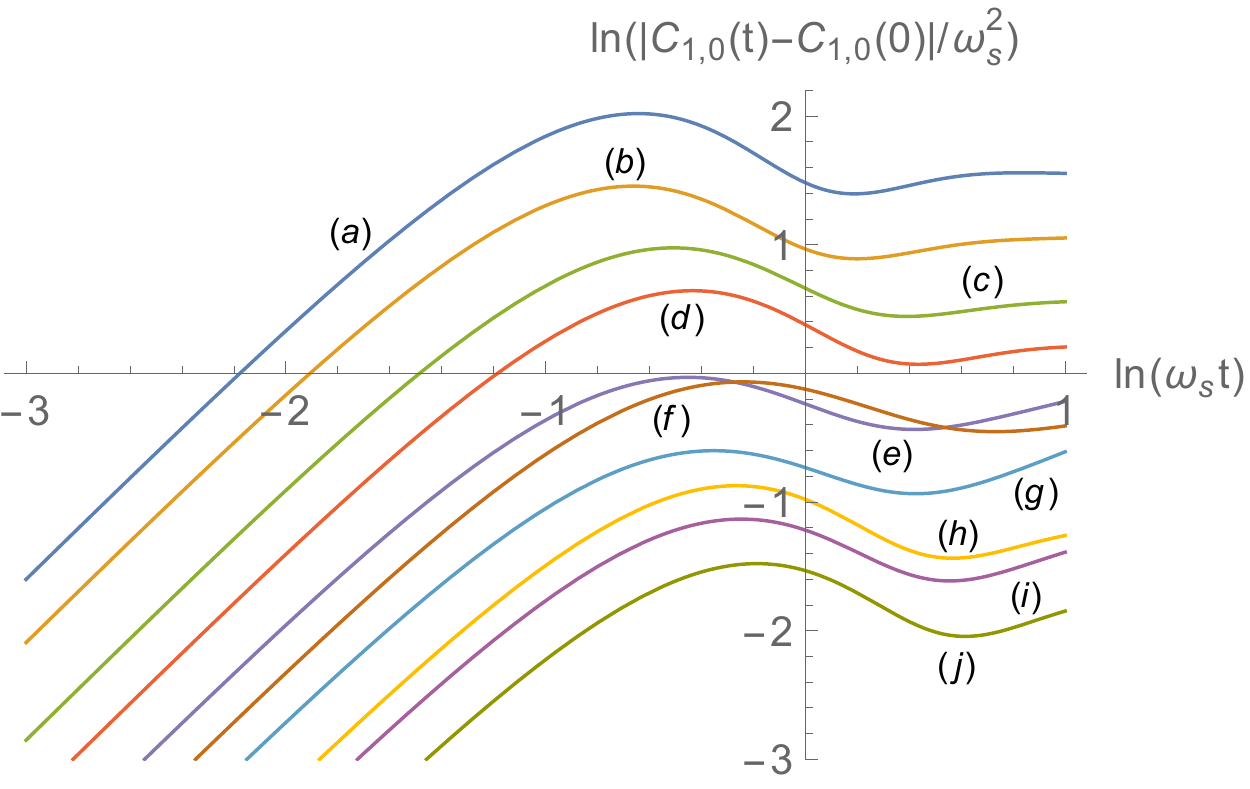}
\vspace{0.2cm}
\caption{(Color online) The quantity 
$\ln\left|\left(C_{1,0}(t)-C_{1,0}(0)\right)/\omega_s^2\right| $ versus $\ln\left( \omega_s t\right)$ for $\exp\left\{-3\right\}\leq\omega_s t\leq e$, $J\left(\omega\right)=q\omega_s \left(\omega/\omega_s\right)^{\alpha}\exp\left\{- l \omega/\omega_s \right\}\left|\ln\left(\omega/\omega_s \right)\right|^n$, $q=1$ and different values of the parameters $\alpha$, $l$ and $n$. The curve $(a)$ corresponds to $\alpha=2$, $l=1.2$ and $n=6$; $(b)$ corresponds to $\alpha=1$, $l=1$ and $n=6$, $(c)$ corresponds to $\alpha=1.6$, $l=1.1$ and $n=4$; $(d)$ corresponds to $\alpha=2$, $l=1.3$ and $n=4$; $(e)$ corresponds to $\alpha=0.3$, $l=0.9$ and $n=4$; $(f)$ corresponds to $\alpha=1$, $l=1$ and $n=2$; $(g)$ corresponds to $\alpha=0.2$, $l=1$ and $n=4$; $(h)$ corresponds to $\alpha=1.5$, $l=1.5$ and $4$; $(i)$ corresponds to $\alpha=1$, $l=1.4$ and $n=4$, $(j)$ corresponds to $\alpha=1.6$, $l=1.7$ and $n=4$. Over short times each curve tends to an asymptotic line with slope $2$.}
\label{Fig2SD}
\end{figure}

\begin{figure}[t]
\centering
\includegraphics[height=6.25 cm, width=10.25 cm]{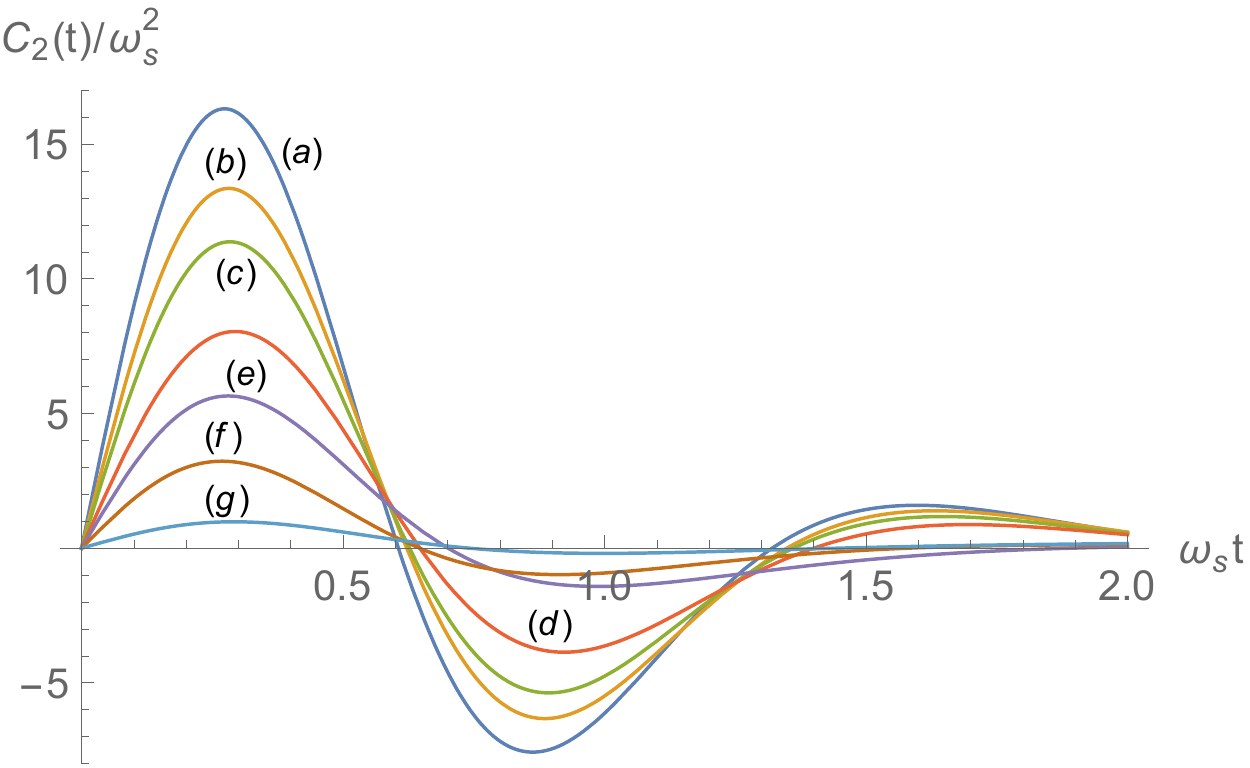}
\vspace{0.2cm}
\caption{(Color online) The quantity 
$C_{2}(t)/\omega_s^2 $ versus $ \omega_s t$ for $0\leq\omega_s t\leq 2$, $J\left(\omega\right)=q\omega_s \left(\omega/\omega_s\right)^{\alpha}\exp\left\{- l \omega/\omega_s \right\}\left|\ln\left(\omega/\omega_s \right)\right|^n$, $q=1$ and different values of the parameters $\alpha$, $l$ and $n$. The curve $(a)$ corresponds to $\alpha=4.8$, $l=1.8$ and $n=6$; $(b)$ corresponds to $\alpha=5$, $l=1.9$ and $n=6$, $(c)$ corresponds to $\alpha=4.9$, $l=1.9$ and $n=6$; $(d)$ corresponds to $\alpha=5$, $l=2$ and $n=6$; $(e)$ corresponds to $\alpha=3$, $l=1.1$ and $n=2$; $(f)$ corresponds to $\alpha=2.1$, $l=1.1$ and $n=4$; $(g)$ corresponds to $\alpha=1$, $l=1$ and $n=4$.}
\label{Fig3SD}
\end{figure}

\begin{figure}[t]
\centering
\includegraphics[height=6.25 cm, width=10.25 cm]{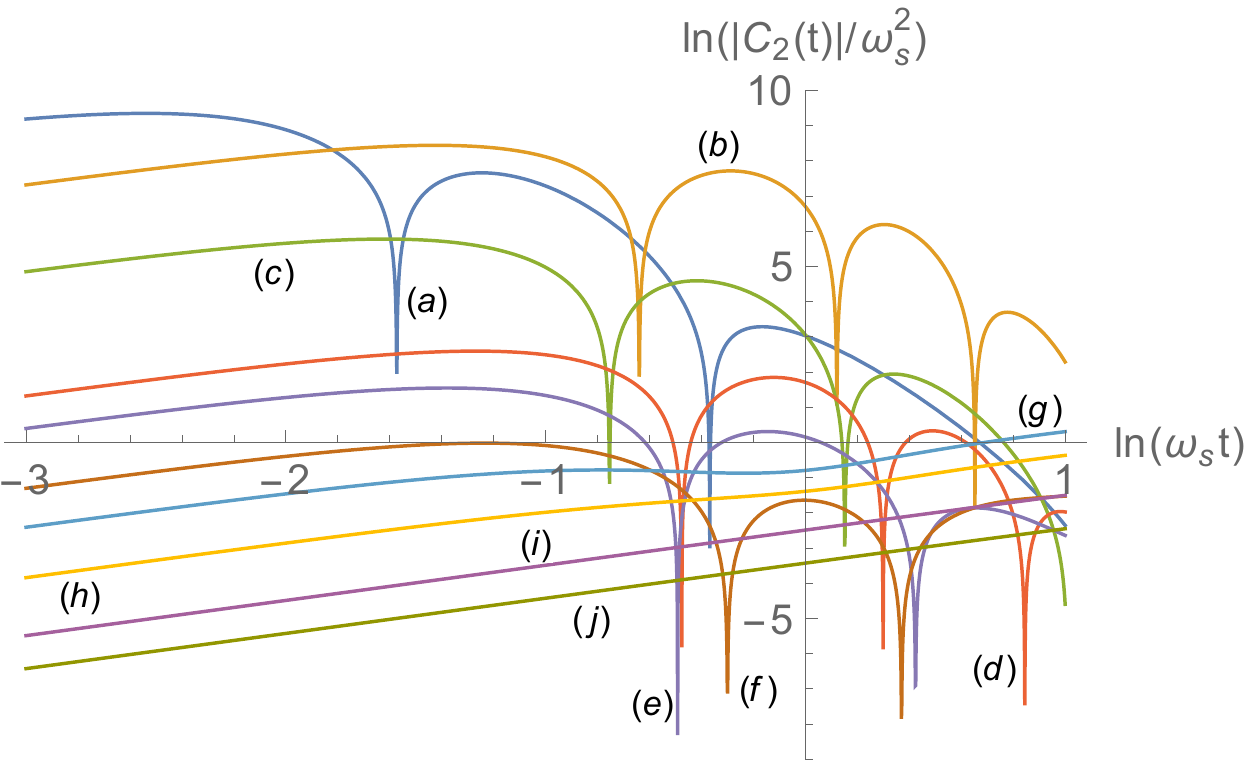}
\vspace{0.2cm}
\caption{(Color online) The quantity 
$\ln\left(\left|C_{2}(t)\right|/\omega_s^2\right) $ versus $\ln\left( \omega_s t\right)$ for $\exp\left\{-3\right\}\leq\omega_s t\leq e$, $J\left(\omega\right)=q\omega_s \left(\omega/\omega_s\right)^{\alpha}\exp\left\{- l \omega/\omega_s \right\}\left|\ln\left(\omega/\omega_s \right)\right|^n$, $q=1$ and different values of the parameters $\alpha$, $l$ and $n$. The curve $(a)$ corresponds to $\alpha=3.4$, $l=0.3$ and $n=2$; $(b)$ corresponds to $\alpha=10$, $l=2$ and $n=2$, $(c)$ corresponds to $\alpha=5$, $l=1$ and $n=2$; $(d)$ corresponds to $\alpha=5$, $l=1.9$ and $n=6$; $(e)$ corresponds to $\alpha=2$, $l=1$ and $n=4$; $(f)$ corresponds to $\alpha=1$, $l=1$ and $n=4$; $(g)$ corresponds to $\alpha=0.1$, $l=1$ and $n=4$; $(h)$ corresponds to $\alpha=1$, $l=1.9$ and $6$; $(i)$ corresponds to $\alpha=1.2$, $l=8$ and $n=6$, $(j)$ corresponds to $\alpha=0.1$, $l=10$ and $n=2$. Over short times each curve tends to an asymptotic line with slope $1$.}
\label{Fig4SD}
\end{figure}

\begin{figure}[t]
\centering
\includegraphics[height=6.25 cm, width=10.25 cm]{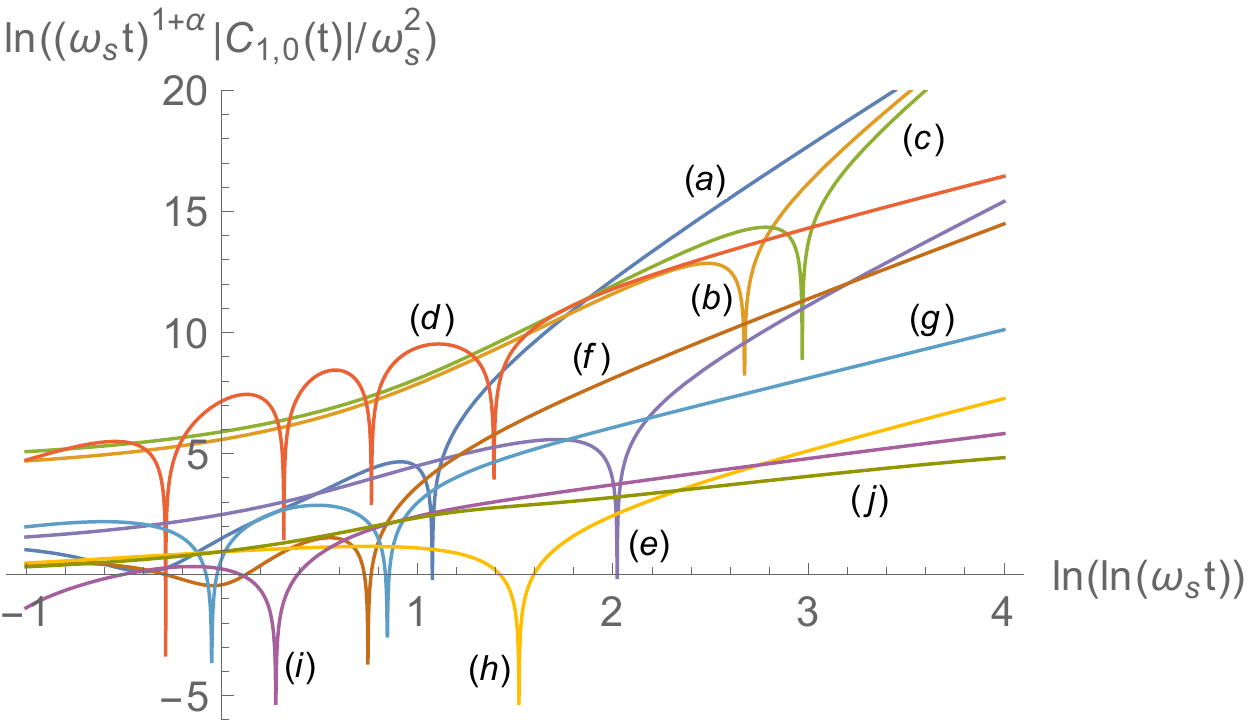}
\vspace{0.2cm}
\caption{(Color online) The quantity 
$\ln\left(\left( \omega_s t\right)^{1+\alpha}\left|C_{0,1}(t)\right|/\omega_s^2\right) $ versus $\ln\left(\ln\left( \omega_s t\right)\right)$ for $\exp\left\{1/e\right\}\leq\omega_s t\leq \exp\left\{ \exp\left\{4\right\}\right\}$, $J\left(\omega\right)=q\omega_s \left(\omega/\omega_s\right)^{\alpha}\exp\left\{- l \omega/\omega_s \right\}\left|\ln\left(\omega/\omega_s \right)\right|^n$, $q=1$ and different values of the parameters $\alpha$, $l$ and $n$. The curve $(a)$ corresponds to $\alpha=2$, $l=1$ and $n=6$; $(b)$ corresponds to $\alpha=0.4$, $l=2$ and $n=6$, $(c)$ corresponds to $\alpha=0.3$, $l=10$ and $n=6$; $(d)$ corresponds to $\alpha=7$, $l=2$ and $n=2$; $(e)$ corresponds to $\alpha=0.5$, $l=2$ and $n=4$; $(f)$ corresponds to $\alpha=2$, $l=1$ and $n=4$; $(g)$ corresponds to $\alpha=3.5$, $l=1$ and $n=2$; $(h)$ corresponds to $\alpha=0.4$, $l=0.01$ and $2$; $(i)$ corresponds to $\alpha=2$, $l=1$ and $n=2$, $(j)$ corresponds to $\alpha=0.01$, $l=10$ and $n=2$. Over long times each curve tends to an asymptotic line.}
\label{Fig5SD}
\end{figure}

\begin{figure}[t]
\centering
\includegraphics[height=6.25 cm, width=10.25 cm]{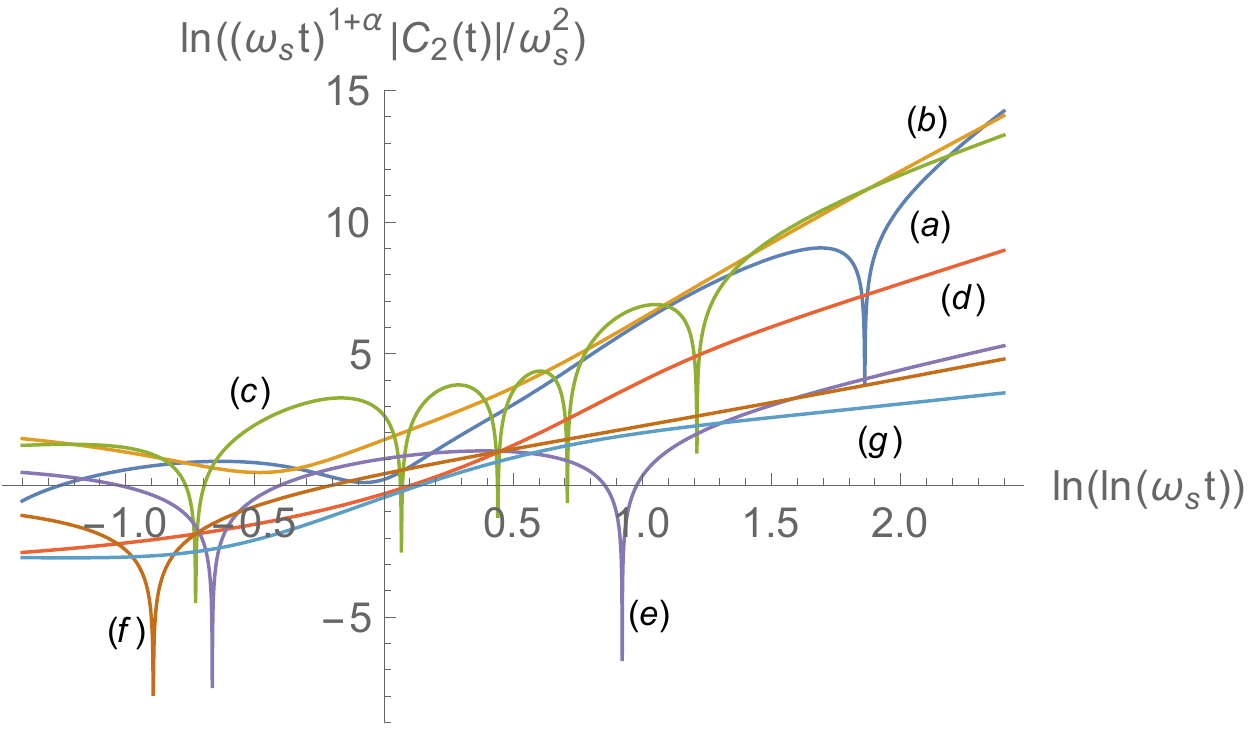}
\vspace{0.2cm}
\caption{(Color online) The quantity 
$\ln\left(\left( \omega_s t\right)^{1+\alpha}\left|C_{2}(t)\right|/\omega_s^2\right) $ versus $\ln\left(\ln\left( \omega_s t\right)\right)$ for $\exp\left\{\exp\left\{-1.4\right\}\right\}\leq\omega_s t\leq \exp\left\{\exp\left\{2.4\right\}\right\}$, $J\left(\omega\right)=q\omega_s \left(\omega/\omega_s\right)^{\alpha}\exp\left\{- l \omega/\omega_s \right\}\left|\ln\left(\omega/\omega_s \right)\right|^n$, $q=1$ and different values of the parameters $\alpha$, $l$ and $n$. The curve $(a)$ corresponds to $\alpha=2$, $l=1$ and $n=6$; $(b)$ corresponds to $\alpha=1$, $l=0.1$ and $n=6$, $(c)$ corresponds to $\alpha=5$, $l=2$ and $n=4$; $(d)$ corresponds to $\alpha=1$, $l=5$ and $n=4$; $(e)$ corresponds to $\alpha=2$, $l=0.5$ and $n=2$; $(f)$ corresponds to $\alpha=0.2$, $l=0.01$ and $n=2$; $(g)$ corresponds to $\alpha=1$, $l=1.5$ and $n=2$. Over long times each curve tends to an asymptotic line.}
\label{Fig6SD}
\end{figure}

Numerical computations concerning the real part of the BCF are displayed in Figures \ref{Fig1SD}, \ref{Fig2SD} and \ref{Fig5SD}. The short time algebraic behavior is suggested by the initial linear growth of the curves plotted in Figure \ref{Fig1SD} and confirmed by the parallel asymptotic lines appearing in the $\log$-$\log$ plot of Figure \ref{Fig2SD}. The long time logarithmic relaxations are confirmed by the asymptotic lines plotted in Figure \ref{Fig5SD}. Numerical computations regarding the imaginary part of the BCF are displayed in Figures \ref{Fig3SD}, \ref{Fig4SD} and \ref{Fig6SD}. The short time linear growth is suggested by the curves plotted in Figure \ref{Fig3SD} and confirmed by the parallel asymptotic lines appearing in the $\log$-$\log$ plot of Figure \ref{Fig4SD}. The long time logarithmic relaxations are confirmed by the asymptotic lines plotted in Figure \ref{Fig6SD}.



\section{Conclusions}\label{5}

The BCF is a fundamental quantity in the analysis of open quantum systems. This function depends on the SD of the system and on the temperature of the thermal bath. Usually, the SDs under study in literature are ohmic-like at low frequencies with exponential cut-off at high frequencies. Ohmic-like SDs can describe an open quantum system that is immersed in a cold environment. This is the case, for example, of an impurity qubit that is trapped in a double-well potential and is immersed in a Bose-Einstein-condensed reservoir \cite{SDeng1,SDengBECM2011}. A detailed analysis of the BCF in terms of the SD is performed in Ref. \cite{RitEisfSDJCP2014}, where a powerful method has been developed to express the BCF a sum of damped harmonic oscillations. This approach applies, in particular, to ohmic-like SDs \cite{RitEisfSDJCP2014}.

In various open quantum systems the SDs depart from the ohmic-like structure. For example, the analysis of bacteriochlorophyll molecules in pigment-protein complexes requires the introduction of SDs of log-normal form \cite{bacterioDSs,RitEisfSDJCP2014}. These SDs are characterized by logarithmic singularities at low frequencies. Recently, logarithmic singularities have been considered in the energy spectrum of unstable quantum systems \cite{LogDecayGiraldi}. The corresponding survival amplitude vanishes and the instantaneous energy tends to the minimum energy of the spectrum according to logarithmic relaxations. In light of the scenario described above, here, we have concentrated on SDs with removable logarithmic singularities at low frequencies. The SDs are arbitrarily shaped at higher frequencies. The singularities consist in arbitrary positive, vanishing, or negative powers of logarithmic functions. These forms are introduced as additional factors in the terms of the power series expansion that describes the low frequency structure of the SDs. In this way, the SDs under study include and continuously depart from the power laws of the ohmic-like condition at low frequencies, and are approximately proportional to the form $\omega_s\left(\omega/\omega_s\right)^{\alpha_0}\left(- \ln\left(\omega/\omega_s\right)\right)^{\beta_0}$ for$\omega \ll \omega_s$. 
We remind that $\omega_s$ represents a typical scale frequency of the system. The low frequency power law profiles are enhanced (reduced) or unchanged by positive (negative) or vanishing values of the logarithmic power $\beta_0$. Over high frequencies, $\omega \gg \omega_s$, the SDs are tailored as follows, $J\left(\omega\right)= O\left(\left(\omega/\omega_s\right)^{-1-\chi_0}\right)$ with $\chi_0>0$.

For the SDs characterized above, we have provided an analytical description of the short and long time behavior of the BCF. Over short times, $t \ll 1/ \omega_s$, a common algebraic behavior is found, $C(t)\sim C_{1,T}(0)-\imath l_2 t-l_{1,T} t^2$, both at zero and arbitrarily non-vanishing temperature, on condition that the SD decays sufficiently fast over high frequencies, $J\left(\omega\right)=O\left(\left(\omega/\omega_s\right)^{-1-\chi_0}\right)$ for $\omega\gg\omega_s$ with $\chi_0>2$. The involved coefficients $l_{1,T}$ and $l_2$ depend on integral properties of the SD. As far as the long time behavior is concerned, at non-vanishing temperatures the real part of the BCF shows regular dependence on the low frequency structure of the SD. In fact, the function $C_{1,T}(t)$ is approximately proportional to the form
$\omega_s T \left(\omega_s t\right)^{-\alpha_0} \ln^{\beta_0}\left(\omega_s t\right)$ for $t \gg 1/\omega_s$, 
if the ohmicity parameters $\alpha_0$ differs from odd natural values. Faster long time relaxations appear for odd natural values of the ohmicity parameter. At zero temperature, the real part of the BCF exhibits regular dependence on the low frequency behavior of the SD. In fact, the function $C_{1,0}(t)$ is approximately proportional to the expression
$\omega_s^2 \left(\omega_s t\right)^{-1-\alpha_0} \ln^{\beta_0}\left(\omega_s t\right)$ for $t \gg 1/\omega_s$, 
on condition that the ohmicity parameter is not an even natural number. Faster long time relaxations are obtained for even natural values of the ohmicity parameter. In the transition from vanishing to arbitrary non-vanishing temperature, the power law term of the dominant long time decay of the real part of the BCF changes from $\left(\omega_s t\right)^{-1-\alpha_0}$ into the slower decay $\left(\omega_s t\right)^{-\alpha_0}$. Consequently, the decay of the whole BCF slows down and becomes arbitrarily slow for vanishing values of the ohmicity parameter, $\alpha_0\to0^+$, and non-negative values of the logarithmic power $\beta_0$. The imaginary part of the BCF shows regular dependence on the low frequency profile of the SD. In fact, the function $C_{2}(t)$ is approximately proportional to the form $\omega_s^2 \left(\omega_s t\right)^{-1-\alpha_0} \ln^{\beta_0}\left(\omega_s t\right)$ for $t \gg 1/\omega_s$, 
if the ohmicity parameter differs from odd natural values. Faster long time relaxations appear for odd natural values of the ohmicity parameter.

The ohmic-like condition is studied as a particular case by making the logarithmic power $\beta_0$ vanish. The short time algebraic behavior of the BCF still holds for ohmic-like SDs with the required sufficiently fast decay, $\chi_0>2$, at high frequencies. Instead, the real and imaginary parts of the BCF vanish over long times according to inverse power laws. In conclusion, a variety of long time relaxations, that are slower than exponential decays and arbitrarily faster or slower than 
inverse power laws, can be interpreted in terms of removable logarithmic singularities in the SDs of open quantum systems that interact with thermal baths.

\appendix
\section{Details} \label{A}

The BCF, given by Eq. (\ref{CT}), is represented in terms of cosine and sine transforms of the SD. The function $C_{1,T}(t)$ vanishes asymptotically due to the Riemann-Lebesgue lemma if the relationship (\ref{cond1T}) is fulfilled, while the function $C_2(t)$ vanishes over long times since the SD is summable.

The short and long time behaviors of the functions $C_{1,T}(t)$ and $C_2(t)$ are studied in the dimensionless variables $\nu$ and $\tau$, that are defined as follows, $\nu=\omega/\omega_s$ and $\tau=\omega_s t$. For the sake of convenience, we introduce the functions $\Xi_{1,T}\left(\tau\right)$ and $\Xi_{2}\left(\tau\right)$ via the following forms, $\Xi_{1,T}\left(\tau\right)= C_{1,T}\left(\tau/\omega_s\right)/\omega_s^2$ and $\Xi_{2}\left(\tau\right)= C_2\left(\tau/\omega_s\right)/ \omega_s^2$. Notice the dependence of the function $\Xi_{1,T}\left(\tau\right)$ on the temperature $T$. According to Eq. (\ref{CT}), the functions $\Xi_{1,T}\left(\tau\right)$ and $\Xi_{2}\left(\tau\right)$ read as below,
\begin{eqnarray}
&&\hspace{-0em}\Xi_{1,T}\left(\tau\right)=
\int_0^{\infty}\Omega_T\left(\nu\right)\cos \left(\nu \tau\right)\, d \nu, \label{Xi1} \\
&&\hspace{-0em}\Xi_{2}\left(\tau\right)=
\int_0^{\infty}\Omega\left(\nu\right)\sin \left(\nu \tau\right)\, d \nu, \label{Xi2}
\end{eqnarray}
where the auxiliary function $\Omega_T\left(\nu\right)$ is defined as $\Omega_T\left(\nu\right)= \Omega\left(\nu\right)
\coth \left(\left(\omega_s \nu\right)/\left(2 T\right)\right)$. The Mellin transforms \cite{BleisteinBook,Wong-BOOK1989} of the functions $\Xi_{1,T}\left(\tau\right)$ and $\Xi_{2}\left(\tau\right)$ are given by the following forms,
\begin{eqnarray}
\hspace{-0em}\hat{\Xi}_{1,T}(s)=\cos\left(\frac{\pi s}{2}\right) \,\Gamma\left(s\right)\hat{\Omega}_{T} (1-s), \label{Xi1s}\\
\hspace{-0em}\hat{\Xi}_{2}(s)=\sin\left(\frac{\pi s}{2}\right) \,\Gamma\left(s\right)\hat{\Omega} (1-s). \label{Xi2s}
\end{eqnarray}
 The fundamental strips \cite{BleisteinBook,Wong-BOOK1989} of the Mellin transforms $\hat{\Xi}_{1,T}(s)$ and $\hat{\Xi}_{2}(s)$ depend on the asymptotic behavior of the auxiliary function $\Omega\left(\nu\right)$ and are discussed below.

The short time behavior of the function $C_{1,T}(t)$ is given by the function $\Xi_{1,T}\left(\tau\right)$ for $\tau\ll 1$. Let the auxiliary function $\Omega\left(\nu\right)$ belong to the first class of SDs introduced in Section \ref{2}. The fundamental strip of the function $\hat{\Xi}_{1,T}(s)$ is $0<\mathrm{Re}\,s<\min\left\{1,\alpha_0\right\}$. 
Let the strip $\mu_0\leq \mathrm{Re}\,s\leq \delta_0$ exist such that the function $\hat{\Omega}_T\left(1-s\right)$, or the meromorphic continuation, vanishes for $T>0$ in the strip as follows,
\begin{eqnarray}
\hspace{-1em}\hat{\Omega}_T\left(1-s\right)= O\left(\left|\mathrm{Im}\, s\right|^{-\zeta_0}\right), \label{cond0OmegasT}
\end{eqnarray}
 for $|\mathrm{Im}\, s|\to+\infty$, where $\zeta_0>1/2$. Let the parameters $\mu_0$ and $\delta_0$ fulfill the following constraints, $\mu_0\in \left(\max\left\{-4,-\chi_0\right\},-2\right)$ with $\chi_0>2$, and $\delta_0\in \left(0,\min\left\{1,\alpha_0\right\}\right)$. Since the following asymptotic relationship \cite{GradRyz} holds for $\left|\mathrm{Im}\, s\right|\to +\infty$,
\begin{eqnarray}
&&\hspace{-4em}\left|\cos\left(\frac{\pi s}{2}\right) \,\Gamma\left(s\right)\right| \sim \left|\sin\left(\frac{\pi s}{2}\right) \,\Gamma\left(s\right)\right|\sim \left(\frac{\pi}{2}\right)^{1/2}\left|\mathrm{Im}\, s\right|^{\mathrm{Re}\,s-1/2}, \label{SimEq1}
\end{eqnarray}
 Eq. (\ref{cond0OmegasT}) induces a sufficiently fast decay of the function $\hat{\Xi}_{1,T}(s)$ as $\left|\mathrm{Im}\, s\right|\to +\infty$, and the singularities of the function 
$\hat{\Xi}_{1,T}(s)$ in $s=0$ and $s=-2$ provide Eq. (\ref{C1Tt0}).
If $T=0$, the fundamental strip of the Mellin transform $\hat{\Xi}_{1,0}(s)$ is $0<\mathrm{Re}\,s<1$. Let the strip $\mu_1\leq \mathrm{Re}\,s\leq\delta_1 $ exist such that the function $\hat{\Omega}\left(1-s\right)$, or the meromorphic continuation, vanishes for $|\mathrm{Im}\, s|\to+\infty$ in the strip as follows,
\begin{eqnarray}
\hspace{-1em}\hat{\Omega}\left(1-s\right)= O\left(\left|\mathrm{Im}\, s\right|^{-\zeta_1}\right), \label{cond1OmegasT}
\end{eqnarray}
 for $|\mathrm{Im}\, s|\to+\infty$, where $\zeta_1>1/2$. Let the parameters $\mu_1$ and $\delta_1$ fulfill the following constraints, $\mu_1\in \left(\max\left\{-4,-\chi_0\right\},-2\right)$ with $\chi_0>2$, and $\delta_1\in \left(0,1\right)$. Under the above conditions the singularities of the function 
$\hat{\Xi}_{1,0}(s)$ in $s=0$ and $s=-2$ provide Eq. (\ref{C1Tt0}), for $T=0$.

We study the short time behavior of the function $C_2(t)$ by analyzing the function $\Xi_{2}\left(\nu\right)$ for $\nu\ll 1$. The fundamental strip of the Mellin transforms $\hat{\Xi}_{2}(s)$ is $\max\left\{-1,-\chi_0\right\}<\mathrm{Re}\,s<1$. 
Let the strip $\mu_2\leq \mathrm{Re}\,s\leq \delta_2$ exist such that the function $\hat{\Omega}\left(1-s\right)$, or the meromorphic continuation, vanishes in the strip as follows,
\begin{eqnarray}
\hspace{-1em}\hat{\Omega}\left(1-s\right)= O\left(\left|\mathrm{Im}\, s\right|^{-\zeta_2}\right), \label{cond0Omegas}
\end{eqnarray}
 for $|\mathrm{Im}\, s|\to+\infty$, where $\zeta_0>1/2$. Let the parameters $\mu_2$ and $\delta_2$ fulfill the following constraints, $\mu_2\in \left(\max\left\{-3,-\chi_0\right\},-1\right)$ with $\chi_0>1$, and $\delta_2\in \left(-1,1\right)$. Under the above conditions the singularity of the function 
$\hat{\Xi}_{2}(s)$ in $s=-1$ provide Eq. (\ref{C2t0}).

The long time behavior of the function $C_{1,T}(t)$ is obtained from the asymptotic expansion of the function $\Xi_{1,T}\left(\tau\right)$ for $\tau \gg 1$. Consider SDs that belong to the first class under study. At non-vanishing temperature, $T>0$, let the strip $\mu_3\leq \mathrm{Re}\,s\leq \delta_3 $ exist such that the function $\hat{\Omega}_T\left(1-s\right)$, or the meromorphic continuation, vanishes in the strip as follows,
\begin{eqnarray}
\hspace{-1em}\hat{\Omega}_T\left(1-s\right)= O\left(\left|\mathrm{Im}\, s\right|^{-\zeta_3}\right), \label{cond0OmegaTs}
\end{eqnarray}
 for $|\mathrm{Im}\, s|\to+\infty$, where $\zeta_3>1/2+ \delta_3$. Let the parameters $\mu_3$ and $\delta_3$ fulfill the following constraints, $\mu_3\in \left(0,\min\left\{1, \alpha_0 \right\}\right)$ and $\delta_3 \in \left(\alpha_{k_3},\alpha_{k_4}\right)$. The parameter $\alpha_{k_3}$ coincides with $\alpha_0$, if $\alpha_0$ is not an odd natural number, or if $\alpha_0=1+2m_0$ and $n_0>0$, otherwise $\alpha_{k_3}$ coincides with the parameter $\alpha_{k_0}$ that is defined in Section \ref{3}. The index $k_4$ is the least natural number that is larger than $k_3$ and such that $\alpha_{k_4}$ does not take odd natural values, or such that $\alpha_{k_4}$ is an odd natural number and $n_{k_4}> 0$.
Under the above conditions the singularity of the function $\hat{\Xi}_{1,T}\left(s\right)$ in $s=\alpha_{k_3}$ provides Eqs. (\ref{C1tInfT})-(\ref{C1tInfT1pl}). For the second class of SDs under study the analysis performed in Refs. \cite{WangLinJMAA1978,Wong-BOOK1989} leads to Eq. (\ref{C1tInfTSD2}).

The long time behavior of the function $C_{1,0}(t)$ is provided by the function $\Xi_{1,0}\left(\tau\right)$ for $\tau \gg 1$. Consider SDs that belong to the first class under study. Let the strip $\mu_4\leq\mathrm{Re}\,s\leq\delta_4 $ exist such that the function $\hat{\Omega}\left(1-s\right)$, or the meromorphic continuation, vanishes in the strip as follows,
\begin{eqnarray}
\hspace{-1em}\hat{\Omega}\left(1-s\right)= O\left(\left|\mathrm{Im}\, s\right|^{-\zeta_4}\right), \label{cond0Omegas}
\end{eqnarray}
 for $|\mathrm{Im}\, s|\to+\infty$, where $\zeta_4>1/2+ \delta_4$. Let the parameters $\mu_4$ and $\delta_4$ fulfill the following constraints, $\mu_4\in \left(0, 1\right)$ and $\delta_4 \in \left(1+\alpha_{k_5}, 1+\alpha_{k_6}\right)$. The power $\alpha_{k_5}$
coincides with $\alpha_0$, if $\alpha_0$ is not an even natural number, or if $\alpha_0=2m_1$ and $n_0>0$, otherwise $\alpha_{k_5}$ coincides with the power $\alpha_{k_1}$ that is defined in Section \ref{3}. The index $k_6$ is the least natural number that is larger than $k_5$ and such that $\alpha_{k_6}$ is not an even natural number, or such that $\alpha_{k_6}$ is an even natural number and $n_{k_6}> 0$. Under the above conditions the singularity of the function $\hat{\Xi}_{1,0}\left(s\right)$ in $s=1+\alpha_{k_5}$ provides Eqs. (\ref{C1tInf0})-(\ref{C1tInf0m1pl}). For the second class of SDs under study the analysis performed in Refs. \cite{WangLinJMAA1978,Wong-BOOK1989} leads to Eq. (\ref{C1tInf0SD2}).

The long time behavior of the function $C_{2}(t)$ is given by the function $\Xi_{2}\left(\tau\right)$ for $\tau \gg 1$. Consider SDs that belong to the first class under study. Let the strip $\mu_5\leq \mathrm{Re}\,s\leq \delta_5$ exist such that the function $\hat{\Omega}\left(1-s\right)$, or the meromorphic continuation, vanishes in the strip as follows,
\begin{eqnarray}
\hspace{-1em}\hat{\Omega}\left(1-s\right)= O\left(\left|\mathrm{Im}\, s\right|^{-\zeta_5}\right), \label{cond0Omegas2}
\end{eqnarray}
for $|\mathrm{Im}\, s|\to+\infty$, where $\zeta_5>1/2+ \delta_5$. Let the parameters $\mu_5$ and $\delta_5$ fulfill the following constraints, $\mu_5\in \left(0,1\right)$ and $\delta_5 \in \left(1+\alpha_{k_7},1+\alpha_{k_8}\right)$. The power $\alpha_{k_7}$
coincides with $\alpha_0$, if $\alpha_0$ is not an odd natural number, or if $\alpha_0=1+2 m_2$ and $n_0>0$, otherwise $\alpha_{k_7}$ coincides with the power $\alpha_{k_2}$ that is defined in Section \ref{3}. The index $k_8$ is the least natural number that is larger than $k_7$ and such that $\alpha_{k_8}$ does not take odd natural values, or $\alpha_{k_8}$ is an odd natural number and $n_{k_8}> 0$.
 Under the above conditions the singularity of the function $\hat{\Xi}_{2}\left(s\right)$ in $s=1+\alpha_{k_7}$ provides Eqs. (\ref{C2tInf0})-(\ref{C2tInfm2pl}). 
For the second class of SDs under study the analysis performed in Refs. \cite{WangLinJMAA1978,Wong-BOOK1989} leads to Eq. (\ref{C2tInf0SD2}). This concludes the demonstration of the present results.

\end{document}